\documentclass{moriond}

\def\Journal#1#2#3#4{{#1} {\bf #2}, #3 (#4)}

\def\PLB{{\em Phys. Lett.}  B}
\def\PRL{\em Phys. Rev. Lett.}

\usepackage{amsmath}

\def\be{\begin{equation}}
\def\ee{\end{equation}}
\def\bea{\begin{eqnarray}}
\def\eea{\end{eqnarray}}

\newcommand{\Photo}{}

\begin{document}
\vspace*{4cm}
\title{Searches for Lepton Flavour Violation at ATLAS and CMS}

\author{Holly Pacey, on behalf of the ATLAS and CMS Collaborations}

\address{Department of Physics, Keble Road, Oxford OX1 3RH, England}

\maketitle
\abstracts{Lepton flavour violation (LFV), and lepton flavour university violation (LFUV), are striking signatures of beyond the Standard Model (BSM) physics.
Recent searches for these at the ATLAS and CMS experiments are presented, using proton-proton collisions with a centre of mass energy of 13 TeV.
A range of models and signatures are considered, including leptoquarks, heavy neutral leptons, LFV in $\tau$ lepton decays, and new measurements of $R(K)$ and $R(J/\Psi)$.
}

\section{Introduction}

Neither lepton flavour nor lepton flavour universality are protected by a Standard Model (SM) symmetry,
so it is `accidental' that they have thus far been generally observed to be conserved.
Whilst neutrino oscillations provide evidence for neutral LFV, charged LFV is heavily suppressed in the SM, for example by the GIM mechanism, to below levels observable at the LHC.
As such, new observations of these would be a clear sign of BSM physics.
Observation of LFUV would also provide evidence of BSM physics.
Recent LFV/LFUV searches from the ATLAS~\cite{ATLAS} and CMS~\cite{CMS} experiments are summarised, using proton-proton collisions at $\sqrt{s}=13$ TeV.

\section{Leptoquarks}

Leptoquark ($LQ$) models connect quarks and leptons, and are often found in Grand Unification Theories.
They manifest as a $LQ$ particle (scalar or vector) which directly couples quarks to leptons, enhancing LFUV processes at the LHC.
$LQ$s can couple to charged or neutral leptons, and the parameter $\beta$ encapsulates the relative branching fractions of these: exclusive coupling to charged (neutral) leptons have $\beta=1$ ($\beta=0$).
Furthermore, in the Yang-Mills scenario, $LQ$s can also couple to gluons, which enhances the LHC production cross section.

\begin{figure}
	\begin{minipage}{0.37\linewidth}
		\centerline{\includegraphics[width=\linewidth]{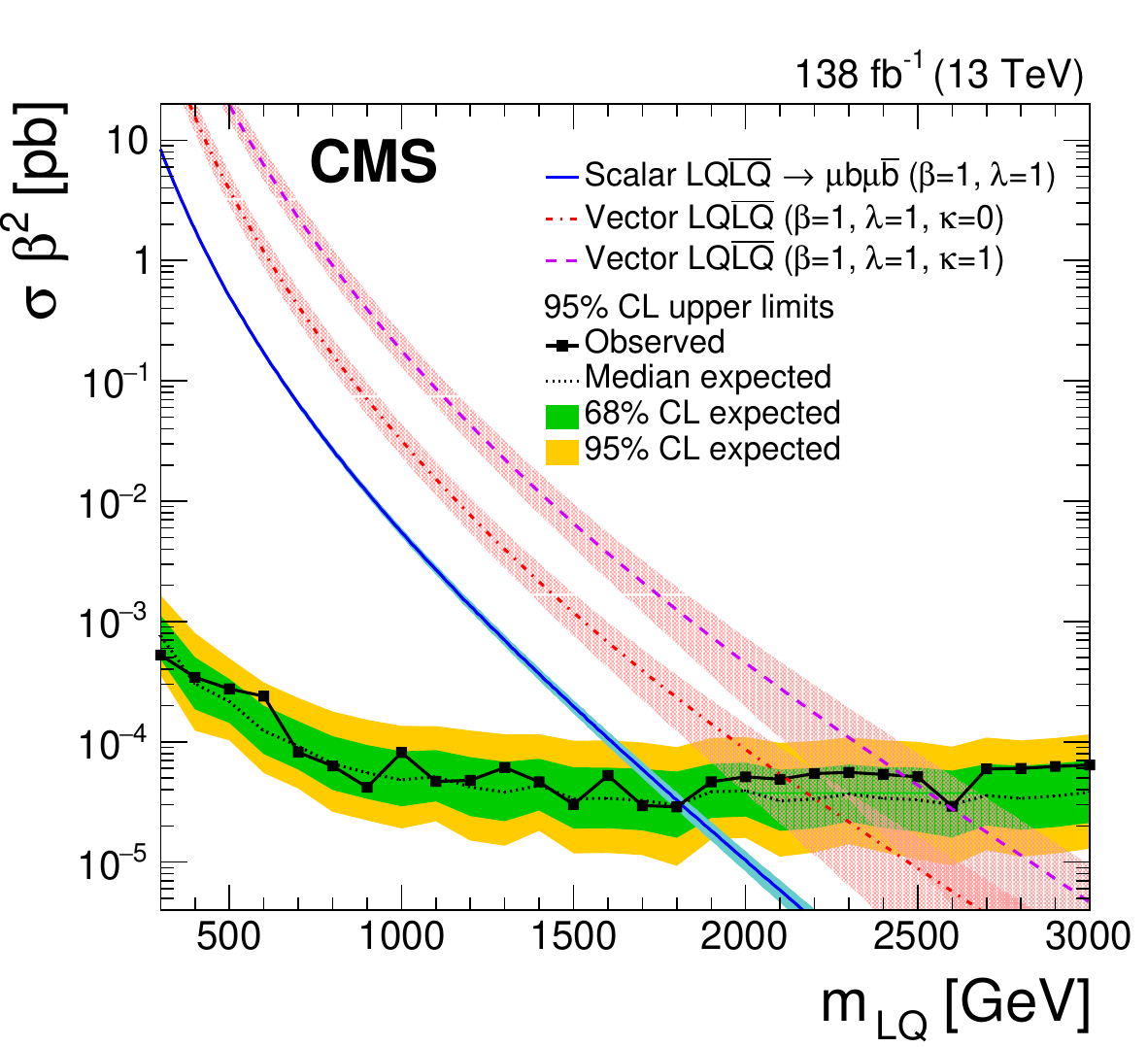}}
	\end{minipage}
	\begin{minipage}{0.44\linewidth}
		\centerline{\includegraphics[width=\linewidth]{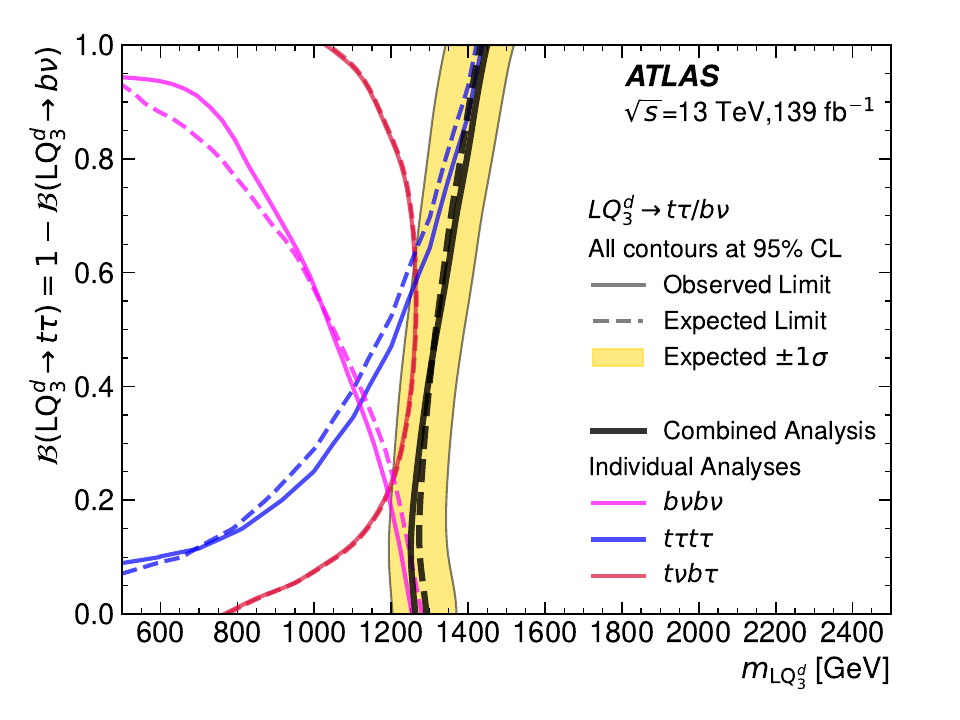}}
	\end{minipage}
	\hfill
	\begin{minipage}{0.44\linewidth}
		\centerline{\includegraphics[width=\linewidth]{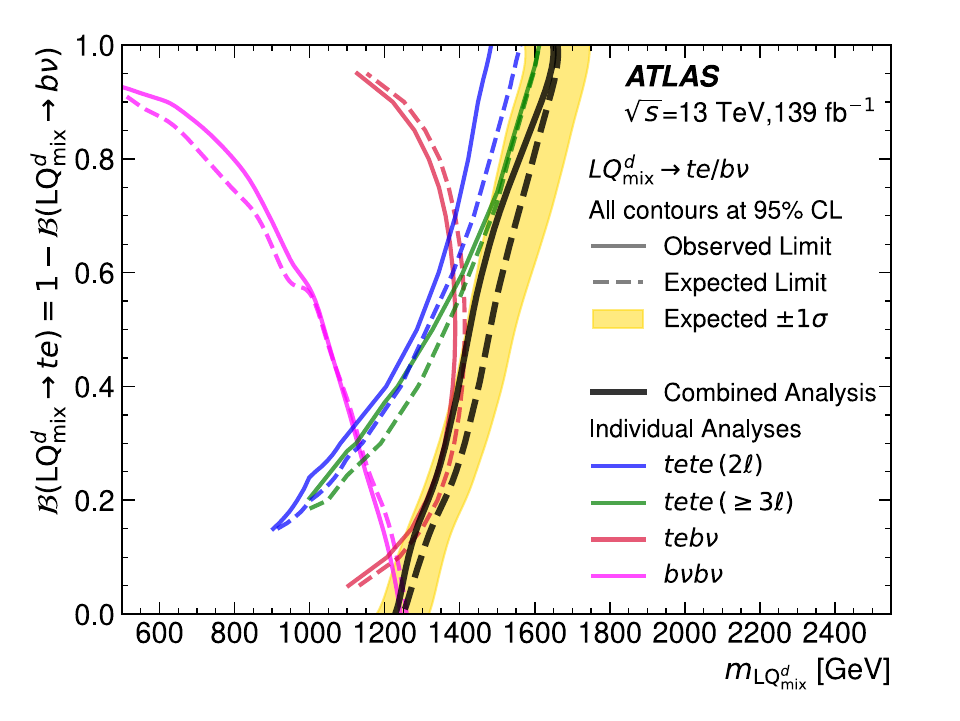}}
	\end{minipage}
	\hfill
	\begin{minipage}{0.44\linewidth}
		\centerline{\includegraphics[width=\linewidth]{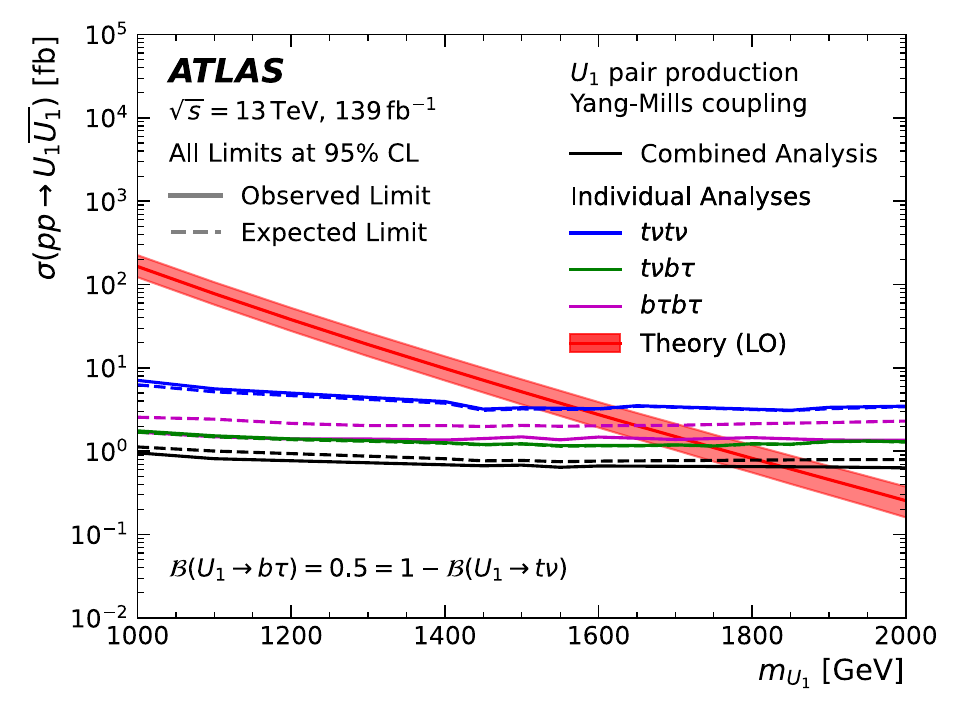}}
	\end{minipage}
	\caption[]{
		Upper limits: on the product of $LQ$ cross section and branching fraction versus mass (top left) from the CMS search~\cite{LQcms}; 
		on the $\beta$ versus LQ mass for $LQ_3^{u/d}$ (top right) or $LQ_{\text{mix}}^{u/d}$ (bottom left) from the ATLAS combination; on the cross section of the vector $LQ$ as a function of mass (bottom right) from the ATLAS combination~\cite{LQatlas}.}
	\label{fig:LQ}
\end{figure}

The latest full Run-2 $LQ$ search from CMS~\cite{LQcms} probes scalar or vector $LQ$s which are pair-produced and couple to $b$ quarks and $\mu$s, leaving a $b\mu b\mu$ final state.
The main SM backgrounds come from Drell-Yan and $t\bar{t}$ processes.
These are estimated using Monte Carlo (MC) simulations, normalised to data using dedicated, process-enriched, high statistics and low signal-contamination Control Regions, with similar kinematics to the search regions.
To isolate the $LQ$ signals, Boosted Decision Trees (BDTs) are trained to classify signal and SM for each $LQ$ mass hypothesis.
A cut is placed on the resulting BDT score distribution, and no significant disagreement between data and the SM background is found.
Hence, the most stringent exclusion limits to date are placed on this model, as shown for example in Figure~\ref{fig:LQ}.
For all scenarios considered, these results improve on the most recent ATLAS limits~\cite{LQbmubmuATLAS}.

In ATLAS, recent focus has been placed on strengthening the exclusion power of existing searches.
A statistical combination of 9 ATLAS searches was performed~\cite{LQatlas} (6 $LQ$ searches and 3 searches originally targetting Supersymmetry), interpreted in terms of a range of models where scalar or vector $LQ$s are pair-produced and couple to 3rd generation quarks.
Either the scalar $LQ$ couples fully to 3rd generation quarks and leptons ($LQ_3^{u/d}$) or the lepton can be 1st/2nd generation ($LQ_{\text{mix}}^{u/d}$).
Limits are placed on $LQ$ masses and $\beta$, with some examples shown in Figure~\ref{fig:LQ}.
Beyond the complementarity between individual search constraints, the combination improves exclusion particularly around $\beta=0.5$, and throughout the vector $LQ$ mass range.
Overall, the most stringent limits to date were placed on the majority of models considered. 

\section{Heavy Neutral Leptons}

Another possible BSM source of LFV appears when considering models to explain tiny SM $\nu$ masses.
See-Saw mechanisms are a common solution, and lead to additional heavy BSM neutrinos ($N$s).
There is no reason to assume that interactions of these heavy neutral leptons would conserve lepton flavour.

\begin{figure}
	\begin{minipage}{0.25\linewidth}
		\centerline{\includegraphics[width=\linewidth]{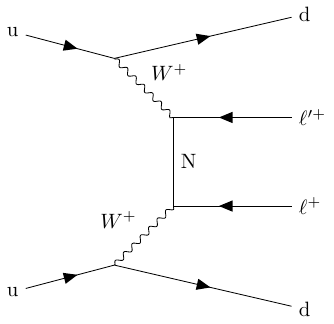}}
	\end{minipage}
	\hfill
	\begin{minipage}{0.39\linewidth}
		\centerline{\includegraphics[width=\linewidth]{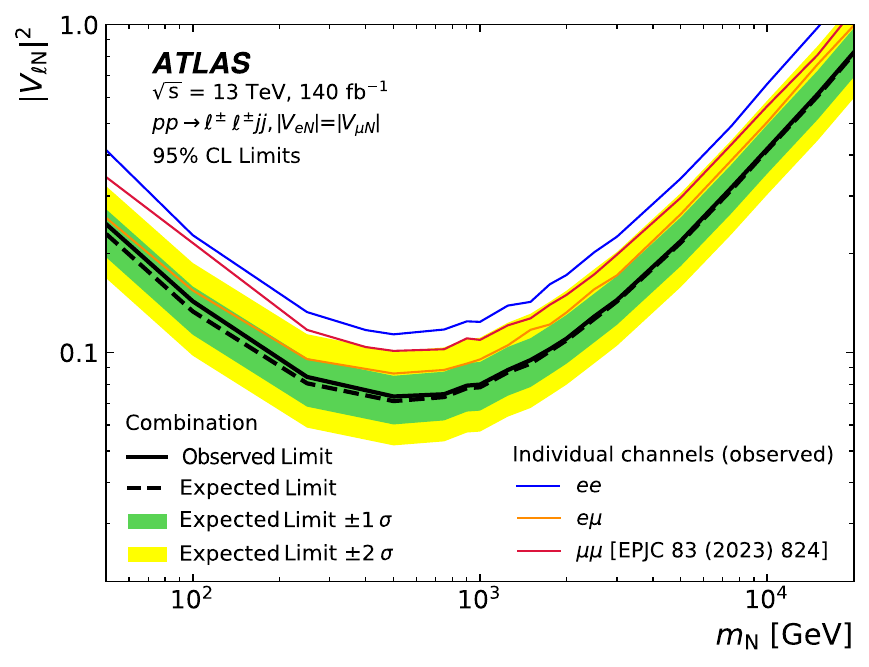}}
	\end{minipage}
	\hfill
	\begin{minipage}{0.33\linewidth}
		\centerline{\includegraphics[width=\linewidth]{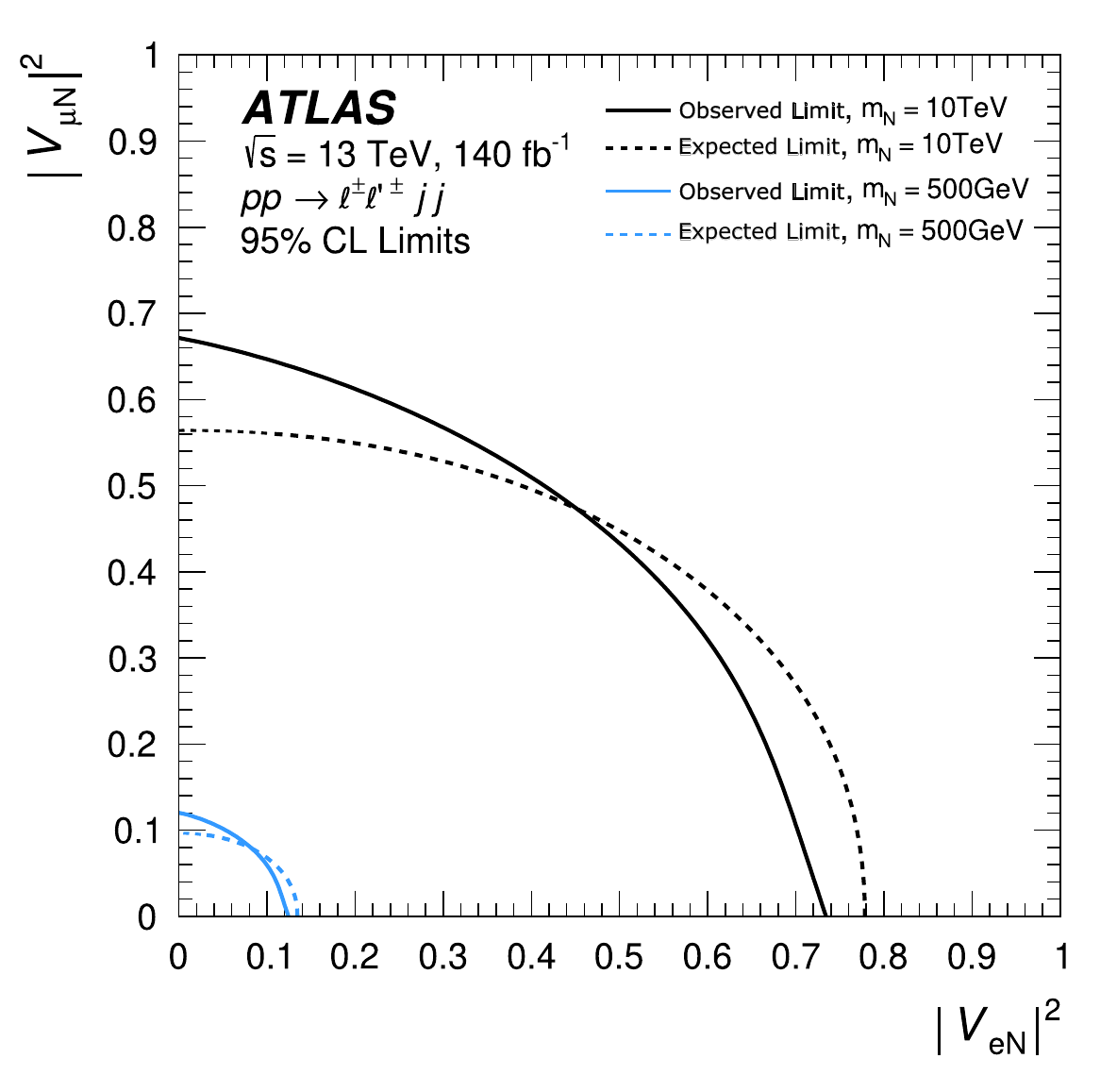}}
	\end{minipage}
	\caption[]{$N$ production Feynman diagram (left), upper limits on $N$ coupling versus mass (middle), and simultaneous constraints on the $e$ and $\mu$ couplings (right) from the ATLAS $N$ search~\cite{HNLatlas}.}
	\label{fig:HNLatlas}
\end{figure}

ATLAS has recently searched for heavy Majorana neutrinos~\cite{HNLatlas}, in a high-energy analogy to neutrinoless double-$\beta$ decay.
The $N$ is produced in $W$-boson scattering, as shown in Figure~\ref{fig:HNLatlas}, leading to a final state of 2 same-charge leptons ($l^{\pm}l^{\pm}$) and 2 forward jets.
The search considers, for the first time, a lepton flavour conserving case where $N$ couples to $e$, and the LFV case where it couples to both $e$ and $\mu$ ($l^{\pm}l^{\pm}=e^{\pm}\mu^{\pm}$).
This follows a similar previous search where $N$ couples only to $\mu$~\cite{HNLmumu}.
A cut-based event selection is used, binning the final fit in the 2nd highest lepton $p_{\text{T}}$.
The dominant VBF $W^{\pm}W^{\pm}/WZ$ backgrounds are estimated using the Control Region approach.
Fully data-driven estimates are used for backgrounds containing fake leptons or electrons with a misidentified charge.
No significant deviation is found between data and the SM background, beyond a $1\sigma$ deficit of data in the (most sensitive to the signal) $p_{\text{T}}>80$ GeV bin in the $e\mu$ channel.
Exclusion limits are placed on the $N$ mass and couplings ($|V_{lN}|$), as seen in Figure~\ref{fig:HNLatlas} for the scenario where the couplings to electrons and muons are equal.
Here the $e\mu$ channel has the best sensitivity due to the deficit.
A statistical combination of the 3 channels considered in this and the previous paper was performed, which improves the limits and allows simultaneous constraint of the $e$ and $\mu$ couplings.

\section{$B$ and $\tau$ physics}

Recent CMS results have also explored more model-agnostic tests for LFV and LFUV, through measurements of branching fraction ratios or searches for enhancements of rare LFV processes.
Firstly, CMS tested lepton flavour universality between muons and taus in $B_c^+$ decays, in a measurement of $R(J/\Psi)$~\cite{BCcms} in 2018 data, as defined in Eq.~\ref{eq:RJpsi}.
\begin{equation}
	R(J/\Psi) = \frac{Br(B_c^+ \rightarrow (J/\Psi \rightarrow \mu^+\mu^-)\tau^+\nu_{\tau})}{Br(B_c^+ \rightarrow (J/\Psi \rightarrow \mu^+\mu^-)\mu^+\nu_{\mu})}
\label{eq:RJpsi}
\end{equation}
The result follows a measurement from LHCb~\cite{BcLHCb} which saw a $2\sigma$ excess above the SM prediction (0.2582).
The analysis considers the case where the $\tau$ decays to a $\mu$, such that both the numerator and denominator have a $3\mu$ final state.
Complex kinematic variables are used to distinguish the these to define two event selections,
The main background, $\pi$s or $K$s faking $\mu$s, is reduced using a Neural Network classifier cut, and estimated using a data-driven template with a normalisation determined in the fit.
Two kinematic distributions are fitted simultaneously to extract the final result, given in Eq.~\ref{eq:RJpsiResult}, which is $0.3\sigma$ from the SM prediction.
\begin{equation}
	R(J/\Psi) = 0.17_{-0.17}^{+0.18} (\text{stat.})_{-0.22}^{+0.21} (\text{syst.})_{-0.18}^{+0.19} (\text{theo.}) = 0.17 \pm 0.33
	\label{eq:RJpsiResult}
\end{equation}

\begin{figure}
	\begin{minipage}{0.44\linewidth}
		\centerline{\includegraphics[width=\linewidth]{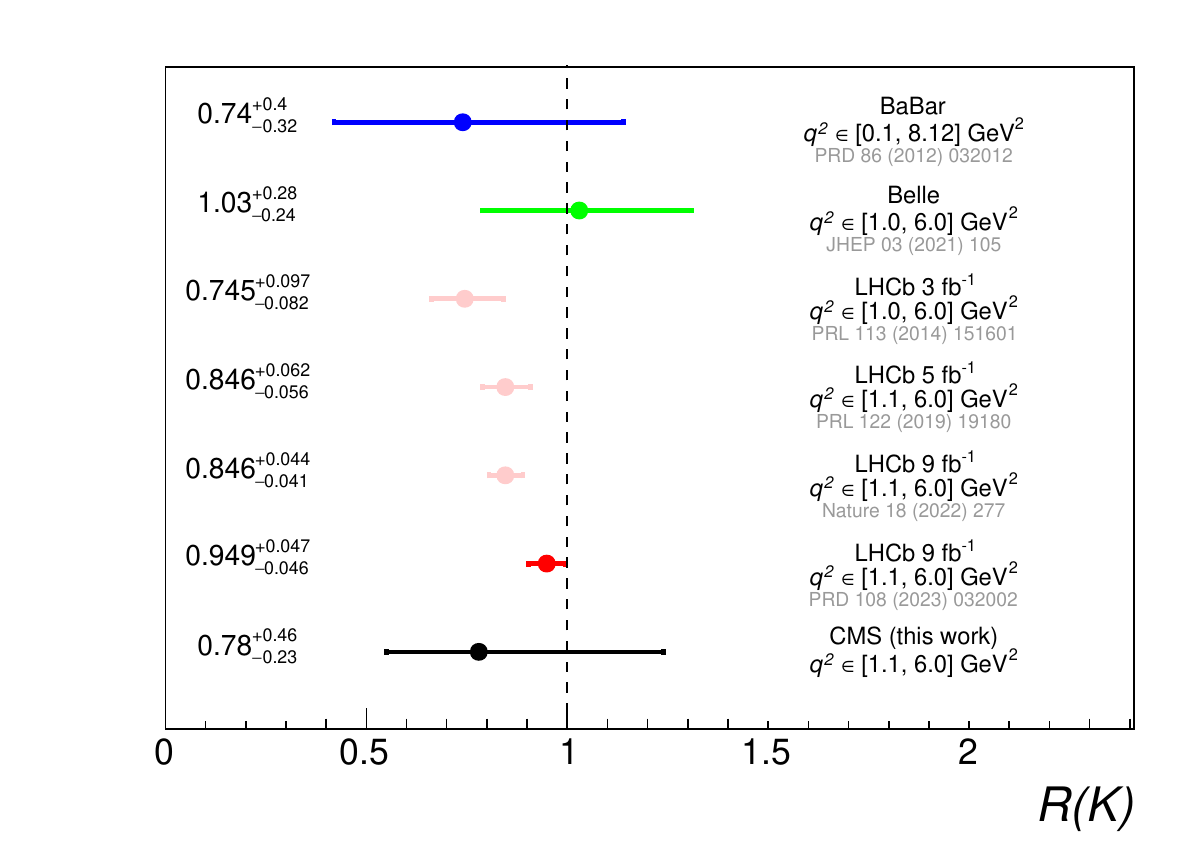}}
	\end{minipage}
	\hfill
	\begin{minipage}{0.44\linewidth}
		\centerline{\includegraphics[width=\linewidth]{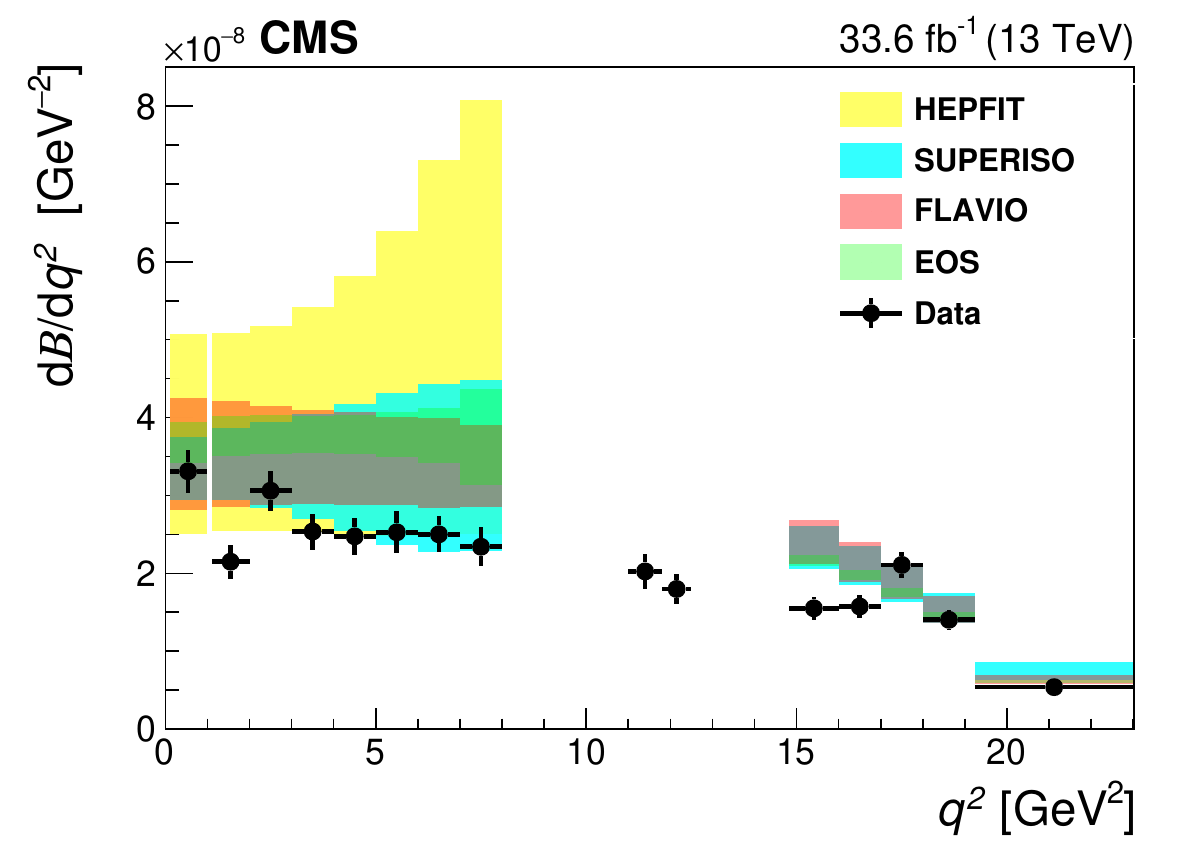}}
	\end{minipage}
	\caption[]{CMS's $R(K)$ measurement~\cite{RKcms} compared to other results (left) and the different measurement (right).}
	\label{fig:RKcmsLFVTAUcms}
\end{figure}

Secondly, CMS performed a measurement of $R(K)$~\cite{RKcms}, defined in Eq.~\ref{eq:RK}.
The measurement uses the same $1.1<q^2<6.0$ GeV$^2$ range as $R(K)$ results from the LHCb and Belle experiments.
\begin{equation}
	R(K) = \frac{Br(B^{\pm} \rightarrow K^{\pm}\mu^+\mu^-)}{Br(B^{\pm} \rightarrow K^{\pm}e^+e^-)}
	\Big/ \frac{Br(B^{\pm} \rightarrow K^{\pm}(J/\Psi \rightarrow \mu^+\mu^-))}{Br(B^{\pm} \rightarrow K^{\pm}(J/\Psi \rightarrow e^+e^-))}
	\label{eq:RK}
\end{equation}
The 2018 `B-parking' data-stream is used, passing a tag+probe trigger that provides 10 billion unbiased $b$-hadron decays.
To reconstruct electrons, the usual CMS electron ID algorithm is uniquely retrained to access electron candidates down to a $p_{\text{T}}$ of 1 GeV.
Most backgrounds are estimated using MC, whilst the combinatorial background is taken from an exponential function.
To select the signal, BDT classifiers are trained, and a cut placed on the score.
After this cut, the $M(B)=M(K\mu\mu)$ distribution is fitted in each channel.
The dominant uncertainties come from limited statistics, and the background parametrisation choice.
The result is shown in Figure~\ref{fig:RKcmsLFVTAUcms}, and is consistent with the SM prediction and results from other experiments.
In this figure the differential measurement is also shown; the predictions slightly overestimate the data measurement, which is consistent with results from LHCb~\cite{KmumuLHCb}. 

Finally, CMS searched for the LFV $\tau\rightarrow3\mu$ decay~\cite{LFVTAUcms}.
This occurs via neutrino mixing in the SM so is heavily suppressed ($B(\tau\rightarrow3\mu)\sim 10^{-55}$), and will only be visible at the LHC with BSM.
The search uses 2017--2018 data, and combines results with a previous result using 2016 data~\cite{LFVTAUcmsOLD} to improve sensitivity.
Signals are searched for as resonances in the $M(\mu\mu\mu)$ spectrum, considering several categories of mass resolution and BDT classifier score.
Two $\tau$ production channels are used.
The first and most sensitive channel, heavy flavour decays $D_s^+\rightarrow \tau^+ \nu_{\tau}$ and $B\rightarrow\tau+X$, have a larger signal yield with lower $p_{\text{T}}$, but also a larger background.
The signal yield is normalised to that of $D_s^+\rightarrow \pi^+\phi \rightarrow \mu\mu\mu$ to reduce the dependence on cross-section predictions and trigger/selection efficiencies.
The second, $W$ decays, have a smaller signal focused at higher $p_{\text{T}}$, but also a smaller background.
All categories and channels are fitted simultaneously, and the results are statistically limited.
The branching ratio limit is $B(\tau\rightarrow3\mu) < 2.9\times10^{-8}$ at 90\% CL, which is the best hadron collider result to date.

\section{Conclusion}

Recent results for LFV and LFUV from the ATLAS and CMS collaborations was presented.
Whilst no signs of BSM physics were found, improved constraints have been placed on $LQ$ and $N$ models.
Furthermore, the first measurements of $R(K)$ and $R(J/\Psi)$ at CMS were summarised.

{\def\thefootnote{}\footnotetext{\textcopyright~Copyright [2018] CERN for the benefit of the ATLAS and CMS Collaborations. Reproduction of this article or parts of it is allowed as specified in the CC-BY-4.0 license}}

\end{document}